\documentclass[mathleft]{an}
\usepackage{graphicx}
\usepackage{times}
\usepackage{natbib}
\bibpunct{(}{)}{;}{a}{}{,}
\begin{document}

\Pagespan{1}{}
\Yearpublication{2012}%
\Yearsubmission{2012}%
\Month{00}%
\Volume{000}%
\Issue{00}%

\title{CSS091109:035759+102943 --  a candidate polar}

\author{A.D.~Schwope\inst{1}\fnmsep\thanks{Corresponding author:
  \email{aschwope@aip.de}\newline}
\and  B.~Thinius\inst{2}
}
\titlerunning{CSS091109:035759+102943 --  a candidate polar}
\authorrunning{A.D. Schwope \& B. Thinius}
\institute{
Leibniz-Institut f\"ur Astrophysik Potsdam (AIP), An der Sternwarte 16, 
14482 Potsdam, Germany
\and 
Inastars Observatory Potsdam, Hermann-Struve-Str. 10, 14469 Potsdam,
  Germany}

\received{June 25, 2012}
\accepted{July 16, 2012}
\publonline{}


\abstract{%
We report optical time-resolved photometry of the CRTS transient
CSS091109:035759+102943. Pronounced orbital variability
with a 114\,min period, large X-ray variability and the IR to X-ray spectral
energy distribution suggest a classification as a magnetic cataclysmic binary,
a likely AM Herculis star or polar. 
}

\maketitle

\section{Introduction}
Polars are magnetic cataclysmic variables (CVs) harboring a strongly magnetic
white dwarf
which accretes matter from a late-type main-sequence star \citep[for a
  comprehensive overview see ][]{warner95}. Accretion in the strongly magnetic
environment happens initially via free-falling streams which are later
threadened by the magnetic field so that a quasi-radial accretion column in
the vicinity of the magnetic pole(s) arises. These accretion regions are
sources of hard and soft X-rays from the thermal plasma and the heated polar
cap(s). Consequently, many of the $\sim$100 now known polars were identified as
optical counterparts of soft X-ray sources detected in the ROSAT all-sky survey
\citep[RASS, e.g.][]{beuschwo94}. More recently, numerous CVs have been identified
spectroscopically in the SDSS \citep[][and references
  therein]{szkody+11}. Despite the large collecting area of XMM-Newton and
Chandra, only a few objects have been discovered with these observatories due
to the low surface density of the objects. 

An alternative route to discover CVs involves comprehensive photometric
surveys performed with a high cadence like the Catalina Real-time Transient
Survey \citep[CRTS; ][]{drake+09}. The CRTS variable CSS091109:035759+102943 was
suggested to be a magnetic cataclysmic variable
in an eMail notification by the variable star network VSNET. This
preliminary identification rested on significant intra-night variability and
transitions from low to high optical states observed in the CRTS and the
association with a cataloged X-ray source \citep[2XMM J035758.6+102938; ][]{watson+09}.

Here we describe time-resolved follow-up photometry of
CSS091109:035759+102943 (henceforth CSS091109) which, combined 
with archival multiwavelength data, led to the
(almost) unique identification of the new variable.

\section{Observations and analysis}
\subsection{Time-resolved photometry}

\begin{figure}[th]
\resizebox{\hsize}{!}
{\includegraphics[clip=]{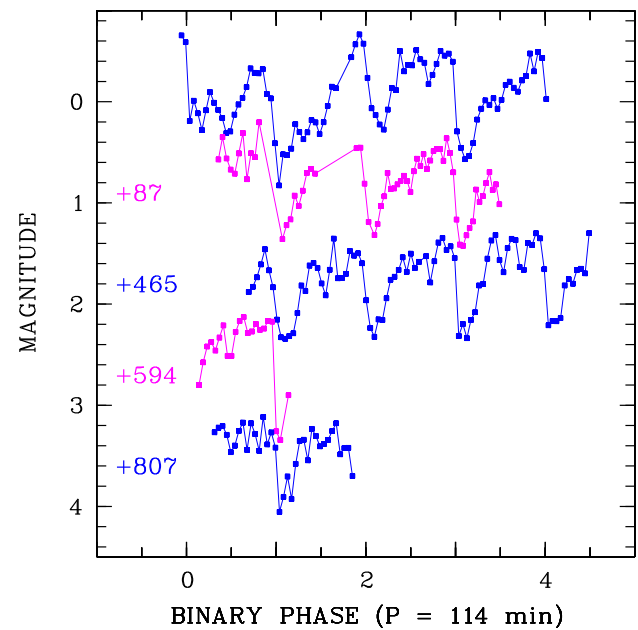}}
\caption{White-light time-resolved photometry of CSS091109 obtained at
  Inastars observatory with a time resolution of 5 min. Shown are data from
  five nights, each vertically shifted by 0.9 magnitudes. 
Data appear in original time sequence, but
transformed to orbital phase using Eq.~\ref{e:eph}.
The phase offset per night is indicated in the figure. }
\label{f:lcori}
\end{figure}

Time-resolved white-light photometry of the field
of the CRTS-variable was obtained during 6 nights 
between Nov.~19, 2009 and Jan.~22, 2010, from Potsdam-Bornim using a
robotically controlled Celestron 
C14 equipped with an ST8XME CCD-camera. The observations were made through a
CLS-filter from Astronomik with full transmission between 450\,nm and 540\,nm
and longward of 635\,nm thus suppressing Hg- and Na-lines from street lamps.

Differential photometric measurements were made with respect to the 
14.5 mag comparison star USNO B1.0 1005-0036054, while USNO B1.0 1005-0036059
(14.5 mag) and USNO B1.0 1005-0036004 (14.7 mag) served as reference stars.
In sum, 331 individual measurements were made with a median 
uncertainty of 0.08\,mag after five minutes of integration. 
Individual zero points displayed nightly variability of about 0.04 mag; both
uncertainties do not affect the result of our study.

We found the object displaying pronounced variability with a
peak-to-peak amplitude of about 0.8 mag. A period search was performed 
using the TSA-package within MIDAS. It revealed just one pronounced peak 
in a Lomb-Scargle periodogram at 114\,min that could already be spotted in the
nightly raw light curves.  

\begin{figure}
\resizebox{\hsize}{!}
{\includegraphics[clip=]{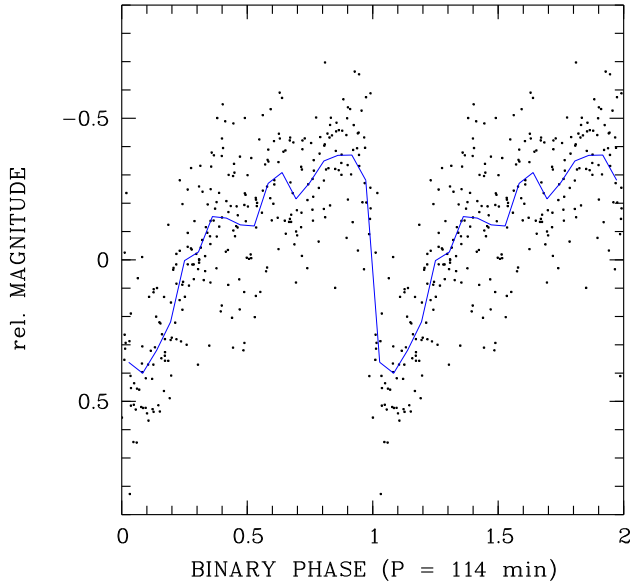}}
\caption{Phase-folded light curve of CSS091109. The mean brightness 
per night was subtracted from the individual data points obtained during that
night. The blue line gives the mean brightness after binning into 18 phase
bins. Phases were computed according to Eq.~\ref{e:eph}.}
\label{f:lcfol}
\end{figure}

The most pronounced
feature in the light curve is a steep decrease of the brightness from a
maximum value into a dip-like minimum. 
We measured the 
times of dip ingress at half light between maximum pre-dip and average
dip-brightness  via graphic cursor in our data (13 epochs with
measurement uncertainty of about 90\,s). A linear regression combining the 13
epochs of this feature revealed an ephemeris 
\begin{equation}
\mbox{BJD}(T_0) = 2455155.3658(4) + E \times 0.079181(1)
\label{e:eph}
\end{equation}
where the numbers in parenthesis indicate the uncertainties in the last
digits. This ephemeris was used to create the final versions of the light
curve plots, both in original time sequence and after phase-binning involving
all individual measurements (Figs.~\ref{f:lcori} and \ref{f:lcfol}). 
The mean light curve displays a saw-tooth shape. 
The dip feature lasts about 0.2 phase units.
After the dip the brightness increases more gradually until the steep 
decrease into the dip phase begins. 
The 114 min periodicity is interpreted as being
due to the orbital motion of a binary star and the apparent brightness of the
star, orbital variability $16.8-18.0$, indicates that it was observed in a high accretion state.

\subsection{Archival CRTS photometry}
\begin{figure}
\resizebox{\hsize}{!}
{\includegraphics[clip=]{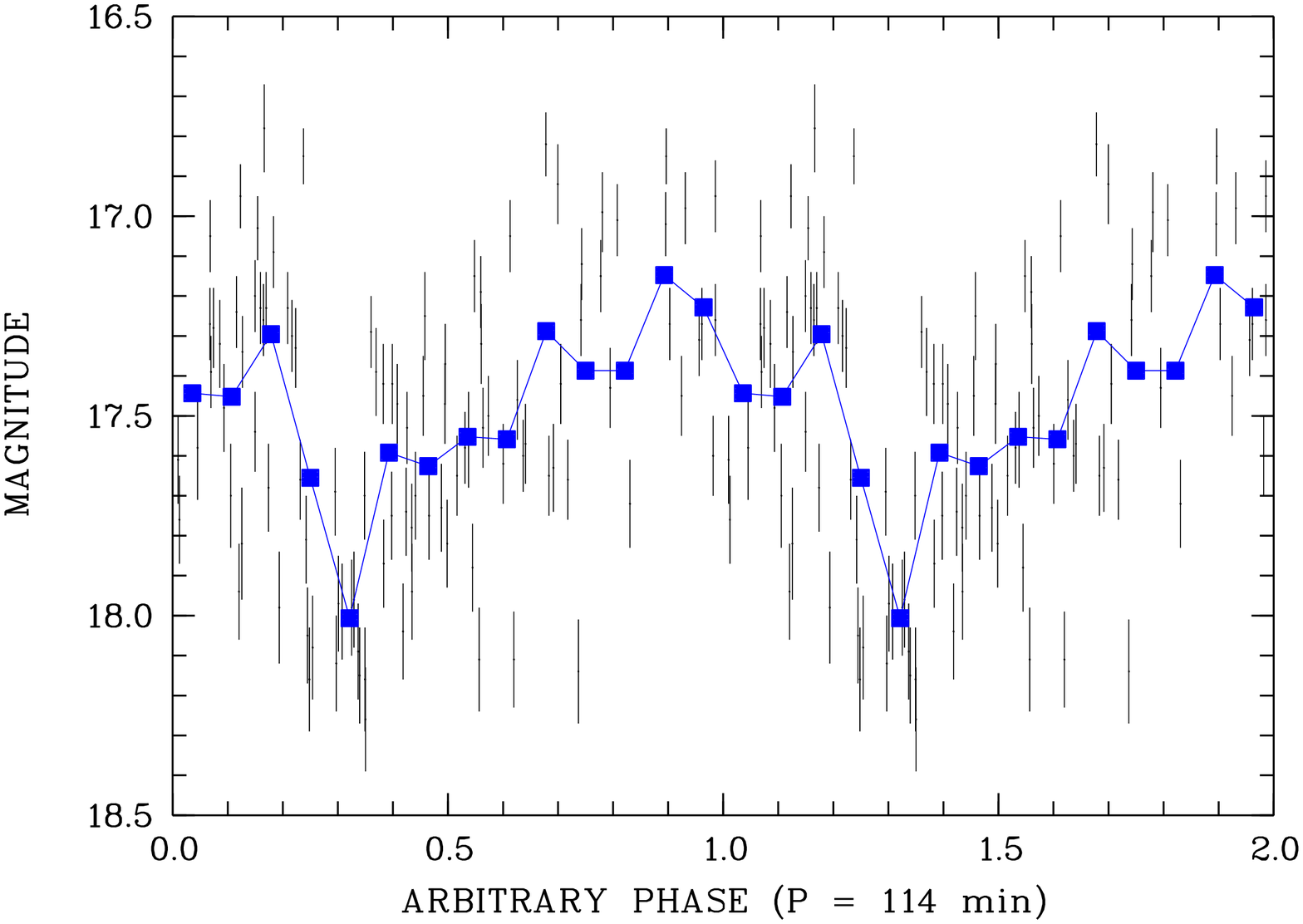}}
\caption{High-state phase-folded light curve of CSS091109 based on CRTS data.
Shown are the original data with their error bars and phase-averaged data
binned into 14 phase bins.}
\label{f:crtsfol}
\end{figure}

We retrieved the currently available CRTS data obtained between Sep.~11, 2007,
and Mar.~16, 2011, from the CRTS archive (211 individual
measurements)\footnote{available at
  http://nesssi.cacr.caltech.edu/catalina/20091109/911091090214133107.html}.  
The data can be grouped into low (orbital variability only below $18.3$ mag),
intermediate (orbital variability between $17.3 - 19.4$) and high accretion
states (orbital variability between $16.8 - 18.2$). An independent
period-search within the high-state data revealed weak evidence for the same
114 min periodicity but could not be used to improve the accuracy of the
ephemeris. A phase-binned representation of the CRTS high-state data using the
period given in Eq.~\ref{e:eph} is reproduced in Fig.~\ref{f:crtsfol}. 
The light curve has similar shape and amplitude of variability as obtained from our
dedicated follow-up. It thus appears that the shape of the light curve
displays a rather stable pattern, at least in the high accretion state.  

\subsection{The spectral energy distribution}
\begin{figure}
\resizebox{\hsize}{!}
{\includegraphics[clip=]{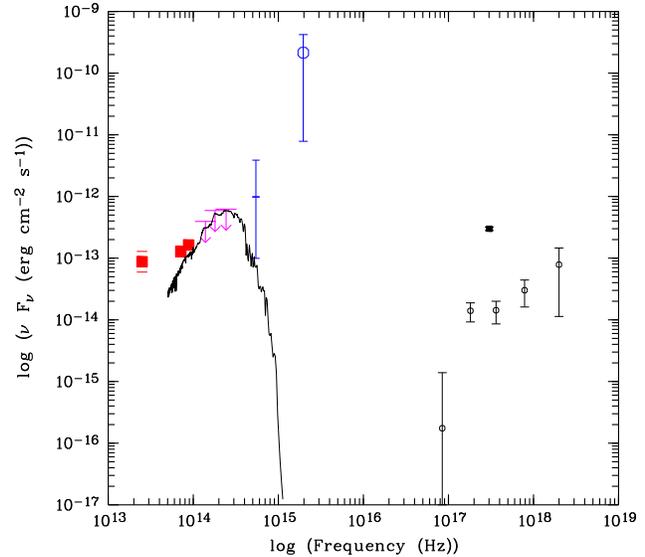}}
\caption{Infrared to X-ray spectral energy distribution of the
  CSS091109. Included are data (from low to high frequencies) from the WISE
  satellite, 2MASS upper limits, the CRTS, GALEX, ROSAT, and XMM-Newton. The
  black curve adapted to the optical/infrared spectral range is a $T_{\rm eff}
  = $3200\,K model spectrum used as template for an assumed M5 donor star.} 
\label{f:sed}
\end{figure}

The VIZIER online tool was used to search for additional observational data
from wide-field imaging surveys. Data are available from WISE, GALEX,
ROSAT-HRI and XMM-Newton. The object has no entry in the 2MASS point source
catalog. We therefore retrieved catalogs from 2MASS in that field and
determined an upper limit magnitude for JHK from 50\%\ completeness. Those
data are indicated with arrows in the $(\nu, \nu f_\nu)$ diagram of
Fig.~\ref{f:sed}. GALEX revealed a detection in the FUV band only 
with a relatively large error (3$\sigma$ detection)
but not in the NUV band.

The photometric data from the Catalina survey are shown in Fig.~\ref{f:sed}
with blue lines.  
The length of the line indicates the amount of orbital variability in the high
and the low accretion states, respectively.

The XMM-Newton count rates from the 2XMM-catalog  were 
transformed to fluxes using the standard energy conversion factors which
assume an AGN-type power law spectrum as emission model. 
CSS091109 was discovered in all five standard energy bands between 0.2 and 10
keV with. XMM-Newton revealed 53 photons during an observation that lasted
15977\,s, i.e.~which covered 2.3 orbital cycles of the 114\,m binary. 
There is a ROSAT discovery as well made with the High Resolution Imager
(HRI). The HRI has no energy resolution, and the count rate of
$5.3\times10^{-4}$\,s$^{-1}$ was transformed to flux assuming a thermal
bremsstrahlung model. The flux observed with ROSAT was found to be a factor
$\sim$10 higher than the flux from the XMM-Newton observation indicating
changes in the mass accretion rate perhaps of that order. The 
XMM-Newton data do not show an obvious indication of a soft component,

WISE detections are reported in the windows W1-W3 at 3.4, 4.2, and 12\,$\mu$m,
respectively but not in W4, the longest wavelength band at 22\,$\mu$m. The
color W1-W2 = 0.61 is slightly redder than expected for any main-sequence star
and in particular redder by about 0.3 mag than a main-sequence star filling
the Roche lobe of a 114 min binary \citep{kirkp+11}. The observed color W2-W3
= 2.56  is even redder than any known brown dwarf. This kind of IR excess
might be due to cyclotron radiation but much more likely due to a circumbinary
disk \citep[see][for similar cases and a more comprehensive
  discussion]{brinkw+07,hoard+07}. 
If we assume that the mass-donating (secondary) star follows the
spectral sequence described by \citet{knigge07} it should have spectral type
dM4.8 which corresponds to $T_{\rm eff} \simeq 3200$\,K. Such a template
spectrum is shown in Fig.~\ref{f:sed}, scaled to make it consistent
with the CRTS photometry, the non-detection in the 2MASS point source catalog,
and to reflect the observed WISE flux in the W1 and W2 bands, respectively. 

The non-detection of CSS091109 in 2MASS, $K_{\rm lim} \geq 15.9$, suggests a
distance larger than 350\,pc if we assume a secondary star with that
brightness following Knigges's donor star sequence, $K=8.08$ at $P_{\rm orb} =
114$\,min.  

\section{Discussion and conclusion}
We interpret the periodicity of 114 min detected in time-resolved photometry
of CSS091109 and covering more than 800 cycles as the orbital period of a
cataclysmic binary. The stable pronounced variability, the change between high
and low brightness states in the optical and at X-ray wavelengths, and the
overall shape of the SED are strongly suggestive of a magnetic binary, a
so-called polar or AM Herculis star. Final confirmation though could be
derived from time-resolved optical spectroscopy or polarimetry.

The object is apparently non-eclipsing which restricts the inclination to
something less than 73\degr. The light curve with its pronounced dip is likely 
shaped by absorption in an accretion stream, irradiation of the stream and the
secondary star, projection effects of the stream and the accretion region on
the white dwarf, and cyclotron beaming. 


X-ray surveys with eROSITA will discover several $10^4$ new compact binaries 
 \citep{predehl+10, as12}. Massive spectroscopic surveys with e.g.~4MOST
 \citep{rdj12} will reveal identification spectra and Gaia the distances for
 most of the CVs. Comprehensive photometric follow-up will be 
needed to uniquely identify the type of a cataclysmic binary and eventually to
uncover the period distribution of the class, an opportunity for  
amateur involvement on a larger scale.


\begin{acknowledgements}
We thank our referee F.V.~Hessman for constructive criticism and 
P.~Hauschildt for providing the PHOENIX model spectrum included in
Fig.~\ref{f:sed}.
\end{acknowledgements}

\bibliographystyle{aa}
\bibliography{css}


\end{document}